\documentclass{article}

\usepackage{amssymb}
\usepackage{graphics}

\begin{document}
\renewcommand{\Bbb}{\mathbb}

\thispagestyle{empty}

%%%%%%%%%%%%%%%%%%%%%%%%%%%%%%%%%%%%%%%%%%%%%%%%%%%%%%%%%%%%%%%%%%%%%%%%%
%                            GREEK                                      %
%%%%%%%%%%%%%%%%%%%%%%%%%%%%%%%%%%%%%%%%%%%%%%%%%%%%%%%%%%%%%%%%%%%%%%%%%
\newcommand{\al}{\alpha}
\newcommand{\bet}{\beta}
\newcommand{\ga}{\gamma}
\newcommand{\del}{\delta}
\newcommand{\ep}{\epsilon}
\newcommand{\epx}{\varepsilon}
\newcommand{\ze}{\zeta}
\renewcommand{\th}{\theta}
\newcommand{\thx}{\vartheta}
\newcommand{\io}{\iota}
\newcommand{\la}{\lambda}
\newcommand{\ka}{\kappa}
\newcommand{\pix}{\varpi}
\newcommand{\rhx}{\varrho}
\newcommand{\si}{\sigma}
\newcommand{\six}{\varsigma}
\newcommand{\yp}{\upsilon}
\newcommand{\om}{\omega}
\newcommand{\phx}{\varphi}
\newcommand{\Ga}{\Gamma}
\newcommand{\De}{\Delta}
\newcommand{\Th}{\Theta}
\newcommand{\La}{\Lambda}
\newcommand{\Si}{\Sigma}
\newcommand{\Yp}{\Upsilon}
\newcommand{\Om}{\Omega}

%%%%%%%%%%%%%%%%%%%%%%%%%%%%%%%%%%%%%%%%%%%%%%%%%%%%%%%%%%%%%%%%%%%%%%%%%
%                         MATH MODE COMMAND                             %
%%%%%%%%%%%%%%%%%%%%%%%%%%%%%%%%%%%%%%%%%%%%%%%%%%%%%%%%%%%%%%%%%%%%%%%%%
\renewcommand{\L}{\cal{L}}
\newcommand{\M}{\ensuremath{\mathcal{M}}}
\newcommand{\N}{\ensuremath{\mathcal{N}}}
\newcommand{\G}{{\cal G}}
\newcommand{\J}{{\cal J}}
\newcommand{\Jb}{\bar{\cal J}}
\newcommand{\be}{\begin{equation}}
\newcommand{\ee}{\end{equation}}
\newcommand{\bea}{\begin{eqnarray}}
\newcommand{\eea}{\end{eqnarray}}
\newcommand{\jt}{\tilde{J}}
\newcommand{\Ra}{\Rightarrow}
\newcommand{\lra}{\longrightarrow}
\newcommand{\ti}{\tilde}
\newcommand{\pj}{\prod J}
\newcommand{\pjt}{\prod\tilde{J}}
\newcommand{\prb}{\prod b}
\newcommand{\prc}{\prod c}
\newcommand{\bft}{|\tilde{\phi}>}
\newcommand{\bfj}{|\phi>}
\newcommand{\lan}{\langle}
\newcommand{\ran}{\rangle}
\newcommand{\bz}{\bar{z}}
\newcommand{\bJ}{\bar{J}}
\newcommand{\tsp}{\tau\hspace{-1mm}+\hspace{-1mm}\si}
\newcommand{\tsm}{\tau\hspace{-1mm}-\hspace{-1mm}\si}
\newcommand{\vacr}{|0\rangle}
\newcommand{\vacl}{\langle 0|}
\newcommand{\IFF}{\Longleftrightarrow}
\newcommand{\phr}{|phys\ran}
\newcommand{\phl}{\lan phys|}
\newcommand{\non}{\nonumber\\}
\newcommand{\tg}{\tilde{g}}
\newcommand{\tM}{\ti{M}}
\newcommand{\hd}{\hat{d}}
\newcommand{\hL}{\hat{L}}
\newcommand{\sir}{\si^\rho}
\newcommand{\mf}{\mathfrak}
\newcommand{\mfg}{{\mathfrak{g}}}
\newcommand{\mfbg}{{\bar{\mathfrak{g}}}}
\newcommand{\mfk}{{\mathfrak{k}}}
\newcommand{\mfc}{{\mathfrak{c}}}
\newcommand{\mfbc}{{\bar{\mathfrak{c}}}}
\newcommand{\mfbk}{{\bar{\mathfrak{k}}}}
\newcommand{\mfn}{{\mathfrak{n}}}
\newcommand{\mfbn}{{\bar{\mathfrak{n}}}}
\newcommand{\mfh}{{\mathfrak{h}}}
\newcommand{\mfbh}{{\bar{\mathfrak{h}}}}
\newcommand{\mfp}{{\mathfrak{p}}}
\newcommand{\mfbp}{{\bar{\mathfrak{p}}}}
\newcommand{\mfU}{{\mf U}}
\newcommand{\p}{\partial}
\newcommand{\pb}{\bar{\hspace{0.2mm}\partial}}
\newcommand{\pl}{\partial_{+}}
\newcommand{\pmi}{\partial_{\hspace{-0.5mm}-}}
\newcommand{\Lg}{\L^{(\mfg})}
\newcommand{\Lgk}{\L^{(\mfg,\mfk)}}
\newcommand{\Lgkp}{\L^{(\mfg,\mfk')}}
\newcommand{\Lk}{\L^{(\mfk)}}
\newcommand{\Lkp}{\L^{(\mfk')}}
\newcommand{\Lc}{\L^{(\mfc)}}
\newcommand{\Lgg}{\L^{(\mfg,\mfg)}(\la)}
\newcommand{\Lgp}{\L^{(\mfg,\mfg')}(\la)}
\newcommand{\Mgp}{\M^{(\mfg,\mfg')}(\la)}
\newcommand{\Mgk}{\M^{(\mfg,\mfk)}(\la)}
\newcommand{\Hgk}{H^{(\mfg,\mfk)}}
\newcommand{\Mgkp}{\M^{(\mfg,\mfk^\prime)}}
\newcommand{\Hgp}{H^{(\mfg,\mfg')}}
\newcommand{\vo}{v_{0\la}}
\newcommand{\Ml}{\M(\la)}
\newcommand{\Ll}{\L(\la)}
\newcommand{\RR}{\mathbb{R}}
\newcommand{\CC}{\mathbb{C}}
\newcommand{\sltwo}{\ensuremath{\mathfrak{sl}_{2}}}
\newcommand{\sltr}{\ensuremath{\mathfrak{sl}(2,\RR)}}
\newcommand{\hst}{\ensuremath{\mathfrak{hs}(2,1)}}
%%%%%%%%%%%%%%%%%%%%%%%%%%%%%%%%%%%%%%%%%%%%%%%%%%%%%%%%%%%%%%%%%%%%%%%%%
%                         MISCELLANEOUS                                 %
%%%%%%%%%%%%%%%%%%%%%%%%%%%%%%%%%%%%%%%%%%%%%%%%%%%%%%%%%%%%%%%%%%%%%%%%%

\newcommand{\e}[1]{\label{e:#1}\end{eqnarray}}
\newcommand{\ind}{\indent}
\newcommand{\noi}{\noindent}
\newcommand{\np}{\newpage}
\newcommand{\hs}{\hspace*}
\newcommand{\vs}{\vspace*}
\newcommand{\nl}{\newline}
\newcommand{\bqu}{\begin{quotation}}
\newcommand{\equ}{\end{quotation}}
\newcommand{\bit}{\begin{itemize}}
\newcommand{\eit}{\end{itemize}}
\newcommand{\ben}{\begin{enumerate}}
\newcommand{\een}{\end{enumerate}}
\newcommand{\ba}{\begin{array}}
\newcommand{\ea}{\end{array}}
\newcommand{\ul}{\underline}
\newcommand{\nn}{\nonumber}
\newcommand{\lef}{\left}
\newcommand{\rig}{\right}
\newcommand{\fra}{\twelvefrakh}
\newcommand{\Bb}{\twelvemsb}
\newcommand{\bT}{\bar{T}(\bz)}
\newcommand{\dagg}{^{\dagger}}
\newcommand{\qd}{\dot{q}}
\newcommand{\cP}{{\cal P}}
\newcommand{\hg}{\hat{g}}
\newcommand{\hh}{\hat{h}}
\newcommand{\hpg}{\hat{g}^\prime}
\newcommand{\htg}{\tilde{\hat{g}}^\prime}
\newcommand{\pri}{\prime}
\newcommand{\bis}{{\prime\prime}}
\newcommand{\lap}{\la^\prime}
\newcommand{\rhop}{\rho^\prime}
\newcommand{\Dgp}{\Delta_{g^\prime}^+}
\newcommand{\Dg}{\Delta_g^+}
\newcommand{\Pro}{\prod_{n=1}^\infty (1-q^n)}
\newcommand{\Pg}{P^+_{\hg}}
\newcommand{\Pgp}{P^+_{\hg\pri}}
\newcommand{\hmu}{\hat{\mu}}
\newcommand{\hnu}{\hat{\nu}}
\newcommand{\hrho}{\hat{\rho}}
\newcommand{\gp}{g^\prime}
\newcommand{\pp}{\prime\prime}
\newcommand{\CM}{\hat{C}(g',M')}
\newcommand{\CI}{\hat{C}(g',M^{\prime (1)})}
\newcommand{\CL}{\hat{C}(g',L')}
\newcommand{\HL}{\hat{H}^p (g',L')}
\newcommand{\HMI}{\hat{H}^{p+1}(g',M^{\prime (1)})}
\newcommand{\da}{\dagger}
\newcommand{\asu}{\ensuremath{\widehat{\hspace{0.5mm}\mathfrak{su}}(1,1)\,}}

\newcommand{\wro}{w^\rho}
\renewcommand{\Box}{\rule{2mm}{2mm}}
\begin{flushright}
July 4, 2002 \\
\end{flushright}
\vs{10mm}

\begin{center}

{\Large{\bf On 7D TFT and 6D chiral CFT}}
\\
\vspace{10 mm}
{\large{Jens Fjelstad\footnote{email: jens.fjelstad@kau.se}}
\vspace{4mm}\\
Department of Physics,\\ Karlstad
University, SE-651 88 Karlstad, Sweden}

\vs{15mm}

{\bf Abstract }
\end{center}

\noindent
We perform a canonical and BRST analysis of a seven-dimensional Chern-Simons theory on a manifold
with boundary. The main result is that the 7D theory induces for consistency a chiral two-form on
the 6D boundary. We also comment on similar behaviour in a five-dimensional Chern-Simons theory
relevant for $\N=4$ supersymmetric Yang-Mills theory in four dimensions.

\np
\setcounter{page}{1}

\section{Introduction}
During the past few years higher dimensional conformal field theories (CFT's) with chiral gauge
fields or, equivalently, (anti) self-dual field strength's, have received considerable attention~
\cite{Witten:comments, Strominger:openp}. The motivation comes primarily from string theory, but
they are undoubtedly interesting enough to study irrespective of their stringy origin. The present
work is mainly relevant for theories with $N$ chiral two-forms in six spacetime dimensions, subject
to an ADE classification. A single two-form describes a free theory while $N>1$ corresponds to
interactions classified by $A_{N}$, $D_{N}$ or, when applicable, $E_{N}$. In the string theory
setting these theories are probably not to be viewed as genuine field theories since they contain
also string-like sources. Still, some aspects are likely to be captured by the field-sector.
Furthermore a new class of genuine field theories were presented in~\cite{Henningson:6dcft}, so
there certainly do exist conformally invariant quantum field theories of chiral two-forms in six
dimensions.
Witten~\cite{Witten:5brane} showed how to interpret the partition function of a chiral two-form in
terms of a holomorphically factorized partition function for a non-chiral two-form. This is very
much in line with how the partition function for a chiral boson in two dimensions, which appears for
instance in the heterotic string, is interpreted. Holomorphic factorization of more general
correlation functions of $2k$-form gauge theories in $4k+2$ dimensions was discussed in~\cite{
HeNiSa:holfac}. A main ingredient in the similarities is that in $4k+2$ dimensions the Hodge $*$
squares to $1$ on a Euclidean manifold~\footnote{Note that on a Lorentzian manifold of the same
dimension $* * =+1$, and it is therefore meaningful to talk about self dual $(2k+1)$-forms.} $X$,
i.e. $* * =-1$, and therefore the Hodge star operator defines a complex structure on the space of
$(2k+1)$-forms in $4k+2$ dimensions.

In~\cite{Verlinde:gdual} it was shown that the Hilbert space of a single chiral two-form is
isomorphic to the Hilbert space of a certain simple topological field theory (TFT) on a class of
compact seven-manifolds without boundary. The action describing this TFT is
\begin{equation}
   \label{7dtft}
   S_7=\La\int_{\M_7}C\wedge dC
\end{equation}
where $C$ is a three-form gauge field. Despite not being related to Chern-Simons forms we will refer
to the corresponding theory as a Chern-Simons theory due to the obvious similarity with three-
dimensional Abelian Chern-Simons theory. The information of $N$ is encoded in $\La$, which for the
$A_N$ theory takes the form $\frac{-N}{4\pi}$. The action (\ref{7dtft}) first appeared in the
literature in~\cite{Schwarz:pnfn} where the partition function on a compact manifold was shown to
yield the Ray-Singer torsion. Canonical quantization of (\ref{7dtft}) was further discussed in~\cite
{Witten:adstft} with applications to the AdS/CFT correspondence.
The relation addressed here is very similar to the relation between three-dimensional TFT and two-
dimensional chiral CFT. In short this relation tells us that the Hilbert space of a TFT is the (
finite dimensional) space of conformal blocks of some rational CFT. Exactly which CFT and which
conformal blocks depends on the specific three-manifold and some additional data.
The conformal blocks are basic building blocks of correlation functions, for instance the zero-point
blocks on the torus are simply the Virasoro characters which are used in constructing the partition
function. It would be very interesting to see precisely how far the analogies to 3D TFT and 2D CFT
can
be drawn, but we also believe there are intrinsic reasons to pursue the relation between 6D chiral
CFT and 7D TFT. On the CFT side, although the free theory is well understood very little is known
about the interacting theory. On the TFT side it would be very interesting from a purely
mathematical point of view to have a new handle on higher-dimensional topology.

The 3D analogue of the problem addressed in this paper, 7D TFT on a manifold with boundary, was
first analysed in~\cite{Witten:qftjones}. It was shown that the boundary turns gauge modes physical
with the dynamics of a chiral WZNW model. The state space then becomes infinite dimensional,
described as a representation of the chiral algebra, and not merely the space of conformal blocks.
This analysis was extended and put more concretely in~\cite{EMSS:canqCS, MS:confzoo}, while in~\cite
{FjHw:99} it was shown using a classical BRST analysis that the relation is a consequence of gauge
invariance. Our main result is the 7D analogue of these results, we show that the theory induces for
consistency a chiral two-form on the 6D boundary. We also show similar behavior in a 5D Chern-Simons
theory, although the analysis is not pursued as far.

The paper is structured as follows. In section~\ref{sec:canonical} we perform a canonical analysis
of the Chern-Simons theory which yields the main result, the classical relation between 7D TFT on a
manifold with boundary and a chiral two-form in 6D. The rationale is to use gauge invariance as a
guiding principle. Starting with the TFT as formulated without boundary we restore the gauge
invariance broken by the boundary through the introduction of new degrees of freedom living on the
boundary. From the extended gauge invariant theory we can gauge fix in different ways proving the
precise relation between the 7D TFT and a 6D chiral two-form. This method has previously been
applied with great success on the non-Abelian Chern-Simons theory in three
dimensions~\cite{FjHw:99}. There are topological subtleties involved in the definition of the TFT,
in particular when discussing quantization. We will, however, assume a setting which is simple
enough to carry out a naive canonical analysis.
Section~\ref{sec:4D} is devoted to a five-dimensional TFT conjectured to play a similar role for
$\N=4$ supersymmetric Yang-Mills theory in four dimensions (which is conformally invariant). Finally
in section~\ref{sec:discussion} we end with a discussion of the results and some speculations on how
to proceed based on analogies with 3D TFT and 2D CFT.

\section{Canonical analysis}
\label{sec:canonical}

We work on a smooth {\em Lorentzian} $7$-manifold $\M_{7}\cong \RR \times \Si$ where
$\p\Si\neq\emptyset$.
Let $$\al, \bet,\ldots \in\{0,\ldots ,5,\perp\}$$ denote tangent indices on $\M_{7}$, $$i,j,\ldots
\in\{1,\ldots ,5,\perp\}$$ on $\Si$ and $$a,b,\ldots\in\{1,\ldots ,5\}$$ on $\p\Si.$
To avoid having to write too many metric determinants, the completely antisymmetric symbols in
various dimensions with indices {\em upstairs} are unconventionally taken to have components
$0,\pm 1$. We have $$\varepsilon^{(d)}_{i_1 \ldots i_d} = g^{(d)}\varepsilon_{(d)}^{i_1 \ldots 1_d}
$$ where $g^{(d)}$ is the determinant of the $d$-dimensional metric. We furthermore define $$
\varepsilon_{(6)}^{ijklmn}=\varepsilon_{(7)}^{0ijklmn}$$ and $$\varepsilon_{(5)}^{abcde}=
\varepsilon_{(6)}^{abcde\perp}.$$ Henceforth the dimensionality will not be written out explicitly,
it is hopefully obvious from the context.
We make frequent use of the symbol
$$\del^{i_1\ldots i_n}_{j_1\ldots j_n}\equiv\del^{i_1}_{[j_1}\del^{i_2}_{j_2}\ldots\del^{i_n}_{j_n]}
$$ with {\em unweighted} anti-symmetrisation, and the relation $$ \varepsilon_{i_1\ldots i_m i_{m+1}
\ldots i_d}\varepsilon^{j_1\ldots j_m i_{m+1}\ldots i_d}=(d-m)!g\del^{j_1\ldots j_m}_{i_1\ldots
i_m}.$$

Consider a three-form gauge field
$C=C_{\al_1\al_2\al_3} dx^{\al_1}\wedge dx^{\al_2}\wedge dx^{\al_3}$ with "dynamics" described by
the action
\begin{eqnarray*}
   S_7 & = & \La \int_{\M_{7}}C\wedge dC\\
   & = & 6\La \int_{\M_7}d^{7}x\varepsilon^{i_1 i_2 \ldots i_6}C_{0 i_1 i_2}\p_{i_3}C_{i_4 i_5 i_6}
   - \La \int_{\M_7}d^{7}x\varepsilon^{i_1 i_2 \ldots i_6}C_{i_1 i_2 i_3}\dot{C}^{(3)}_{i_3 i_4 i_5}
   \\
   & + & 3\La\int_{\M_7}d^{7}x\p_{i_4}\left(\varepsilon^{i_1 i_2 \ldots i_6}C_{i_1 i_2 i_3}C_{0 i_5
   i_6}\right)
\end{eqnarray*}
We will neglect the total derivative term in the forthcoming canonical analysis. This is partly
motivated by the wish for full generality, but since inclusion of this term makes a significant
difference we will comment further on this at the end of this section.

We have the following canonical momenta
\begin{eqnarray}
   \label{constr}
   \pi^{ijk} & = & \La\varepsilon^{ijklmn}C_{lmn}\\
   \label{gausslaw}
   \pi^{0ij} & = & 0
\end{eqnarray}
which are both primary constraints. The canonical equal time Poisson brackets read
\[ \{C_{\al_1 \al_2 \al_3}(x),\pi^{\bet_1 \bet_2 \bet_3}(y)\} = \del^{\bet_1 \bet_2 \bet_3}_{\al_1
\al_2 \al_3}\del^{6}(x-y)\]
and the Dirac bracket corresponding to solving (\ref{constr}) becomes
\begin{equation}
   \label{dirbr}
   \{C_{ijk}(x),C_{lmn}(y)\}^* = -\frac{1}{2\cdot 3!\La g} \varepsilon_{ijklmn}\del^{6}(x-y).
\end{equation}
Henceforth we restrict to the reduced phase space and correspondingly drop the $*$ on the Dirac
bracket.
The canonical Hamiltonian derived from the action above is $$H_0=-6\La\int_{\Si}d^6x \varepsilon^{
ijklmn}C_{0ij}\p_{k}C_{lmn}$$ which gives the equation of motion $$\dot{\pi}^{0ij}=2\cdot3!\La
\varepsilon^{ijklmn}\p_{k}C_{lmn}.$$ This yields the secondary constraint
\begin{equation}
   \label{seconstr}
   G_{ijkl}=0
\end{equation}
where $$G_{ijkl}\equiv \frac{1}{3!}\p_{[i}C_{jkl]}= \p_i C_{jkl} + \p_j C_{ilk} + \p_k C_{ijl} +
\p_l C_{ikj}$$ is the field strength corresponding to the potential $C_{ijk}$.

We would now like to examine the constraint algebra. Define for this purpose the integrated
generators $$G[\la] = \int_{\Si}\la\wedge dC = \int_{\Si}d^6x \varepsilon^{ijklmn}\la_{ij}\p_{k}C_{
lmn} = \frac{1}{4}\int_{\Si}d^6x\varepsilon^{ijklmn}\la_{ij}G_{klmn}$$ where the "test two-form"
$\la$ lie in some class of two-forms which do not vanish at $\p\Si$. Using (\ref{dirbr}) we obtain
the following Poisson bracket
\begin{eqnarray}
   \{G[\la^1],G[\la^2]\} & = & \frac{1}{4^2}\int_{\Si}d^6x d^6y \varepsilon^{ijklmn}\varepsilon^{
   pqrstu}\la^{1}_{ij}(x)\la^{2}_{pq}(y)\{G_{klmn}(x),G_{rstu}(y)\}\nonumber\\
   \label{constrcomm}
   & = & -\frac{3}{\La}\int_{\p\Si}\varepsilon^{abcde}\la^1_{ab}\p_c \la^2_{de}.
\end{eqnarray}
We see that the secondary constraints are not in general first class, although they are not fully
second class either. For instance the constraints $G_{abcd}=0$ are all first class since the
corresponding two-forms have the structure $\la_{a\perp}$. Also if either $\la^1$ or $\la^2$ is
closed, the RHS of (\ref{constrcomm}) vanishes.
We can write the constraint algebra in local form as
\begin{equation}
   \label{localccomm}
   \{ G_{ijkl}(x),G_{mnpq}(y) \} = -\frac{1}{24g\La}\varepsilon_{ijkl[mn}\del^{\perp}_p
   \p^{(y)}_{q]}\del^{6}(x-y)\del(x\in\p\Si)
\end{equation}
where we have included a "boundary delta function" $\del(x\in\p\Si)$. This object was introduced in~
\cite{FjHw:99} and is simply defined by $$\int_{\Si}f(x)\del(x\in\p\Si) = \int_{\p\Si}f(x)|_{x\in\p
\Si}.$$ It is convenient to separate out the first class constraints $G_{abcd}$
\begin{eqnarray}
   \label{1stcl1}
   \{G_{abcd}(x),G_{efgh}(y)\} & = & 0\\
   \label{1stcl2}
   \{G_{abcd}(x),G_{efg\perp}(y)\} & = & 0\\
   \label{2ndcl}
   \{G_{abc\perp}(x),G_{def\perp}(y)\} & = &
   -\frac{1}{24g\La}\left(\varepsilon_{abc[de}\p^{(x)}_{f]}\del^6(x-y)\right)\del(x\in\p\Si).
\end{eqnarray}
Introducing a three-form $H_{abc}$ on $\p\M_7$ with the following equal time Poisson relations
\begin{equation}
   \label{Hpoisson}
   \{H_{abc}(x),H_{def}(y)\} = \frac{1}{24g\La}\varepsilon_{abc[de}\p^{(x)}_{f]}\del^5(x-y)
\end{equation}
it is possible to write first-class functions $$\varphi_{abc} = G_{abc\perp} +
H_{abc}\del(x\in\p\Si).$$ We now consider an "improved" gauge invariant theory specified in terms of
$C_{ijk}$ {\em and} $H_{abc}$, and with the following constraints
\begin{eqnarray}
   \label{fullconstr1}
   \pi^{0ij} & = & 0\\
   \label{fullconstr2}
   G_{abcd} & = & 0\\
   \label{fullconstr3}
   \varphi_{abc} & = & 0.
\end{eqnarray}
Note that this differs from the original theory {\em only} on the boundary, but it still remains to
show exactly how they are related.
As a first step in determining this relation we note that, as was already mentioned, we have
introduced too many extra degrees of freedom. This can also be seen if we try to pick the gauge
$H_{abc}=0$. Since $$ \{\varphi[\la],H_{abc}(x)\}\sim (d\la)_{abc}$$ where $$\varphi[\la] = \int_{
\Si}d^5x\varepsilon^{abcde}\la_{ab}\varphi_{cde}$$ choosing a closed two-form gauge parameter $\la$
(i.e. $d\la =0$) leaves $H_{abc}=0$ invariant. We have therefore introduced more degrees of
freedom than necessary to obtain gauge invariance.

It is relatively simple to write down a BRST charge for the extended theory since there are now only
first class constraints, the only slight difficulty is the fact that the constraints are reducible.
A conventional three-form gauge potential $C$ is associated with two-stage reducibility. If $C$
transforms as $C\rightarrow C+d\La^{(2)}$ then the first reducibility identity is (pictorially)
$\La^{(2)}\sim\La^{(2)} + d\La^{(1)}$ for an arbitrary one-form $\La^{(1)}$, and the second is
$\La^{(1)}\sim \La^{(1)} + d\La^{(0)}$.
In our case there is only one reducibility condition due to the boundary degrees of freedom, and in
terms of the canonical constraints it reads $$\varepsilon^{abcde}\p_{a}G_{bcde}=0.$$
With this data the simplest form of the BRST charge is $$ \Om = \int_{\Si}d^6x\left( \eta^{abcd}G_{
abcd} + \eta^{abc}G_{abc\perp} + \pi^{0ij}\bar{b}_{ij} + \tilde{\eta}\varepsilon^{abcde}\p_{a}P_{
bcde}\right) + \int_{\p\Si}d^5x\eta^{abc}H_{abc}$$ where ($\eta^{abcd},P_{abcd}$), (
$\eta^{abc},P_{abc}$) and ($\bar{b}_{ij},\bar{c}^{ij}$) are conjugate fermionic ghost pairs
such that
and $$gh\, {\eta} = gh\, {\bar{b}} = -gh\, {P} = -gh\, {\bar{c}}=1$$ while
($\tilde{\eta},\tilde{P}$) are bosonic conjugate ghosts-for-ghosts such that $$gh\, {\tilde{\eta}}=-
gh\, {\tilde{P}}=2.$$
We will now pick a gauge fixing fermion to determine the dynamics of $H_{abc}$, choose $$\chi= \frac
{1}{4}\int_{\Si}\left(\varepsilon^{ijklmn}P_{ijkl}C_{0mn} +\dot{C}_{0ij}\bar{c}^{ij} + \sqrt{g}^{-1}
\varepsilon^{abcde}C_{abc}\bar{c}_{de}\right) + \frac{1}{4}\int_{\p\Si}\sqrt{g}^{-1}\bar{c}_{ab}
\left(H_{cde}\varepsilon^{abcde} - \al\pi^{0ab}\right)$$
which implies the following BRST invariant Hamiltonian
\begin{eqnarray}
   H & = & \{\Om,\chi\}\nonumber\\
   & = & 3!\int_{\Si}\varepsilon^{abcdij}G_{abcd}C_{0ij}+\frac{3}{2}\int_{\Si}\varepsilon^{abcde}
   \varphi_{abc}C_{0de} + \frac{1}{2}\int_{\Si}\varepsilon^{ijklmn}P_{ijkl}\bar{b}_{mn}\nonumber\\
   & + & \frac{1}{2}\int_{\Si}\pi^{0ij}\dot{C}_{0ij} + \frac{1}{2}\int_{\p\Si}\sqrt{g}^{-1}\pi^{0}_{
   ab}\left(\varepsilon^{abcde}H_{cde}-\al\pi^{0ab}\right).
\end{eqnarray}
The equations of motion for $C_{0ab}$ read $$\varepsilon^{abcde}H_{cde}\del(x\in\p\Si) = 2\al\pi^{0
ab}\del(x\in\p\Si)$$ and inserting this into the BRST invariant Hamiltonian yields
\begin{eqnarray}
   H & = & H_{\p} + H_{bulk}\nonumber\\
   \label{bdryHam}
   H_\p & = & \frac{3}{2\al}\int_{\p\Si}\sqrt{g}H_{abc}H^{abc}.
\end{eqnarray}
Assuming that the bulk topology is simple enough we can eliminate all bulk degrees of freedom, and
all we have left is the boundary part $H_\p$.
Consider for a moment a three-form $X=dB$ on $\p\M_7$. Starting from the free action $$\int_{\p\M_7}
X\wedge *X$$ for $B$, the Hamiltonian takes the familiar Maxwellian form of a sum over electric and
magnetic field strengths $$H_{X} = \int_{\p\Si}d^5x\sqrt{g}\left(\dot{B}_{ab}\dot{B}^{ab} + X_{abc}X
^{abc}\right).$$
If $X=*X$ the electric components $\dot{B}_{ab}$ are completely determined in terms of the magnetic
$X_{abc}$, so all degrees of freedom are contained in the latter components. Restricting the
Hamiltonian to a self-dual $X$ then gives $$H_{chiral} \sim \int_{\p\Si}d^5x\sqrt{g}X_{abc}X^{abc}$$
which agrees with our Hamiltonian (\ref{bdryHam}) for $H_{abc}$.
Decomposing a free three-form $X=X^++X^-$ one can derive the following commutation relations~\cite{
Henningson:comrels}
\begin{eqnarray*}
   \{X^\pm_{abc}(x),X^\mp_{def}(y)\} & = & 0\\
   \{X^\pm_{abc}(x),X^\pm_{def}(y)\} & = & \pm\frac{1}{4g}\varepsilon_{def[ab}\p_{c]}\del^5(x-y).
\end{eqnarray*}
Observe that the latter relations are (up to an irrelevant numerical factor) identical to the
canonical Poisson relations (\ref{Hpoisson}).
These facts are enough to draw the conclusion that the dynamics of $H_{abc}$ is exactly that of a (
anti) chiral two-form gauge field in six dimensions.
Recall now that we introduced too many degrees of freedom with $H_{abc}$, and that transformations
with closed two-form parameters are in a sense redundant. With our newfound interpretation of
$H_{abc}$ in terms of a chiral two-form, this redundancy just reflects the reducibility of gauge
transformations of the two-form. In the canonical picture this is simply a consequence of $dH=0$,
and the redundant degrees of freedom thus corresponds to the co-exact part of $H_{abc}$. The
dynamics of $H$ then kills these degrees of freedom.

Let us make a couple of final remarks on the canonical analysis.
In the beginning we dropped a boundary term from the constraints. In fact had we included that term,
the constraints would have been first-class. However, the constraints (\ref{seconstr}) would then
have been written
\begin{eqnarray*}
   G_{abc\perp} - \p_{\perp}C_{abc} & = & 0\\
   G_{abcd} & = & 0
\end{eqnarray*}
and the canonical analysis
had been based on a theory different from the one defined by $$dC=0$$ in the bulk. Our rationale was
instead to start out with the conventional TFT and then trying to make sense of this on a
manifold with boundary.

A matter which remains unclear concerns the status of the extended constraints $\varphi_{abc}$ (\ref
{fullconstr3}). If we assume smooth field configurations these constraints tell us that $H_{abc}$ is
set to zero. We believe, however, that this is not the correct way to interpret the constraints.
They can be interpreted differently if we allow for configurations which have $\del(x\in\p\Si)$
support. Unfortunately this interpretation puts the validity of the canonical analysis at risk, more
precisely it is the gauge fixing on the boundary which becomes more difficult to analyse. We have a
heuristic argument for this interpretation based on analogy with three-dimensional Chern-Simons
theory. The analysis performed in this section can equally well be carried out in such a setting,
resulting in a chiral boson on the boundary. In quantum Hall systems this gives rise to so called
edge currents, electric currents which have in fact been experimentally observed. We therefore
propose that $H_{abc}$ are to be thought of as genuine currents on $\p\M_7$, and that $G_{abc\perp}$
is allowed to have $\del(x\in\p\Si)$ support.
Let us just stress once more that this is presently a matter of interpretation and not a settled
issue.

\section{Relation to four dimensions}
\label{sec:4D}
Let us now comment on the relation of the seven dimensional theory to theories in lower dimensions,
in particular to a five-dimensional TFT described by
\begin{equation}
   \label{5dtft}
   S_5 = \ka\int_{\M_5}B^{(NS)}\wedge dB^{(RR)}
\end{equation}
which is presumably related to $\N=4$ SYM in four dimensions in a similar
manner~\cite{Witten:adstft}. The fields $B^{(NS)}$ and $B^{(RR)}$ are two independent two-form gauge
fields which, when the theory occurs in type IIB supergravity, are the Neveu-Schwarz and Ramond-
Ramond two-form fields.
Consider the seven-dimensional theory on $\M_7\cong \M_5\times T^2$ and perform a dimensional
reduction on the torus. By the K\"{u}nneth formula we have $$H^3(\M_7,\RR)\cong H^1(\M_5,\RR)\oplus
H^3(\M_5,\RR)\oplus H^2(\M_5,\RR)\oplus H^2(\M_5,\RR)$$ and accordingly the action (\ref{7dtft})
decomposes into
\begin{eqnarray}
   S_7 & \rightarrow & 3\La\int_{\M_5}C\wedge dA + 3\La\int_{\M_5}A\wedge dC\nonumber\\
      \label{dimred}
   & - & 9\La\int_{\M_5}B^{(RR)}\wedge dB^{(NS)} + 9\La\int_{\M_5}B^{(NS)}\wedge dB^{(RR)}
\end{eqnarray}
where $A$ and $C$ are one- and three-forms respectively. The equations of motion in the bulk states
that all fields are closed, and the action for the two-forms is just (\ref{5dtft}) up to a boundary
term. We will henceforth disregard the first two terms and discuss the two-form theory. It is
convenient to combine $B^{(NS)}$ and $B^{(RR)}$ in a complex two-form $B=B^{(NS)}+iB^{(RR)}$ with
the action
\begin{equation}
   \label{5dctft}
   S_5 = \frac{i\ka}{2}\int_{\M_5}B\wedge d B^*.
\end{equation}
The canonical analysis of this action on $\M_5\cong \RR\times Y$ looks in principle the same as for
the seven-dimensional theory, and we get the following canonical brackets
\begin{eqnarray}
   \{B_{AB}(x),B_{CD}(y)\} & = & 0\nonumber\\
   \label{5dcanbr}
   \{B^*_{AB}(x),B^*_{CD}(y)\} & = & 0\\
   \{B_{AB}(x),B^*_{CD}(y)\} & = & -\frac{i}{\ka g}\varepsilon_{ABCD}\del^4(x-y)\nonumber
\end{eqnarray}
where $A,B,\ldots=1,2,3,4$.
We also have the following constraints
\begin{eqnarray}
   \pi^{0A} & = & 0\nonumber\\
   \label{5dconstr}
   W_{ABC} & = & 0
\end{eqnarray}
where $W_{ABC} = \p_A B_{BC} + \p_B B_{CA} + \p_C B_{AB}$, and we have disregarded a boundary term
as in
section~\ref{sec:canonical}.
Evaluating the constraint algebra yields
\begin{eqnarray}
   \{W_{ABC}(x),W_{DEF}(y)\} & = & 0\nonumber\\
   \label{5dconstralg}
   \{W^*_{ABC}(x), W^*_{DEF}(y)\} & = & 0\\
   \{W_{ABC}(x), W^*_{DEF}(y)\} & = & -\frac{i}{2\ka
   g}\varepsilon_{ABC[D}\del^{\perp}_E\p^{(y)}_{G]}\del^4(x-y)\del(x\in\p Y)\nonumber
\end{eqnarray}
where $\perp$ is a coordinate normal to $\p Y$. Denoting coordinates tangent and normal to $\p Y$ as
$i,j,\ldots =1,2,3$ and $\perp = 4$ we can write
\begin{eqnarray}
   \{W_{ijk}(x),W^*_{lmn}(y)\} & = & 0\nonumber\\
   \label{5dcalg2}
   \{W_{ijk}(x),W^*_{lm\perp}(y)\} & = & 0\\
   \{W_{ij\perp}(x),W^*_{kl\perp}(y)\} & = & -\frac{i}{2\ka g}\left(\varepsilon_{ij[k}\p^{(x)}_{l]}
   \del^4(x-y)\right)\del(x\in\p Y).\nonumber
\end{eqnarray}
To obtain first class constraints we introduce new {\em complex} degrees of freedom $F_{ij}$ on
$\p Y$ with the canonical Poisson relations
\begin{equation}
   \label{ffbarcomm}
   \{F_{ij}(x), F^{*}_{kl}(y)\} = \frac{i}{2\ka g}\varepsilon_{ij[k}\p^{(x)}_{l]}\del^{3}(x-y).
\end{equation}
Modify now the constraints according to $$W_{ij\perp}(x)\rightarrow \Phi_{ij}(x) = W_{ij\perp}(x) +
F_{ij}(x)\del(x\in\p Y)$$ which then become first class.

It is easy to show that such $F_{ij}$ determines a (anti) self-dual two-form on $\p\M_5$. Since we
assume that $\p\M_5$ is a four dimensional Lorentzian manifold we know that $**=-1$, and the space
of complex two-forms on $\p\M_5$ is the direct sum of a self-dual and an anti self-dual part
$\La^{(2)\CC}_{\p\M_5}\cong \La^+\oplus\La^-$ such that $$F^{\pm}\in\La^{\pm} \Rightarrow *F^{\pm}=
\pm iF^{\pm}.$$ Consider an arbitrary real two-form $F$ on $\p\M_5$. It can be written $$F=\frac{1}{
2}(F-i*F)+\frac{1}{2}(F+i*F) = F^+ + F^-$$ where $F^\pm\in\La^\pm$ and $(F^\pm)^* = F^\mp$. As in
six dimensions it is enough to consider the spatial components of a self-dual form since they
contain all independent degrees of freedom, i.e. it is enough to consider the components
$F^\pm_{ij}$. If $F$ is derived from the Maxwell action the only non-zero Poisson relation is $$\{F_
{ij}(x),F^{0k}(y)\} = \del^{k}_{[j}\p_{i]}\del^3(x-y)$$ from which we derive
\begin{eqnarray}
   \{F^\pm_{ij}(x),F^\pm_{kl}(y)\} & = & 0\nonumber\\
   \{F^\pm_{ij}(x),F^\mp_{kl}(y)\} & = & \pm\frac{i}{2}\varepsilon_{kl[i}\p_{j]}\del^3(x-y)
\end{eqnarray}
which with $(F^\pm)^*=F^\mp$ is the relation (\ref{ffbarcomm}) up to an irrelevant numerical factor.
We have thus shown that the new degrees of freedom correspond to a complex self-dual two-form field.
We will not derive the BRST charge, suffice it to say that we believe that it is possible in analogy
with the six-dimensional case to show that the boundary two-form is closed and therefore it can
locally be written $F=dA$ where $A$ is a complex Abelian gauge field.

\section{Discussion}
\label{sec:discussion}
Let us first give a short summary of what we have done. By demanding gauge invariance we showed that
the seven-dimensional TFT (\ref{7dtft}) for consistency induces a chiral two-form on the boundary,
and similarly the five-dimensional TFT (\ref{5dtft}) induces a complex self-dual two-form, likely to
be the field strength of a complex Abelian gauge field.
This parallels the development in 3D TFT where it was shown~\cite{Witten:qftjones, EMSS:canqCS} that
three-dimensional Chern-Simons theory on a three-manifold with a spatial boundary induces a chiral
CFT. In the three-dimensional case, however, this is a result which was never pursued very far. The
rigorous definition of a 3D TFT is instead built on the analysis of Chern-Simons theory on
$\M_3\cong \RR\times\Si_2$ where $\p\Si_2=\emptyset$.
We believe that this is a limitation and that a solid understanding of TFT requires also a rigorous
definition of TFT on $\M_3$ such that $\p\Si_2\neq\emptyset$. One might suspect that this would be
something like topological field theory on manifolds with corners, and in fact there are attempts to
define such theories~\cite{KerlerLyuba:3Dcorner}. However, at the moment this is not sufficiently
developed to make practical use of, and the results of this paper therefore seems difficult to apply
directly. Instead we expect that proceeding with seven-dimensional TFT is easiest on seven-manifolds
without spatial boundary. With the purpose of identifying key steps, let us briefly review some
aspects of three-dimensional topological field theory.

Witten~\cite{Witten:qftjones} has shown how the Hilbert space obtained in quantization of 3D Chern-
Simons theory is related to objects in 2D chiral CFT, and the same ideas were investigated more
concretely by Elitzur, Moore, Schwimmer and Seiberg~\cite{EMSS:canqCS, MS:confzoo}. The basic idea
is to consider Chern-Simons theory, with gauge group $G$, on a three-manifold of topology
$\mathbb{R}\times\Si_2$ with the first factor as time. For compact $\Si_2$ without boundary the
Hilbert space, $\mathcal{H}_{\Si_2}$, is in essence a quantization of the moduli space of flat
connections of a principal $G$-bundle $P\rightarrow\Si_2$. This picture is modified somewhat by
including Wilson line observables since we must then allow for a finite set of marked points on
$\Si_2$ where in particular the curvature is singular. It is known that for compact simply connected
$G$ this moduli space is finite dimensional, and Witten argued convincingly that
$\mathcal{H}_{\Si_2}$ can be identified with the space of chiral blocks on $\Si_2$ for a WZNW model
based on $G$.
Matters are in fact quite different when $\Si_2$ has a boundary, it was noted in~\cite{
Witten:qftjones} that the Hilbert space in this case is infinite dimensional. Part of the Hilbert
space can be described as the quantization of $LG/G$ where $LG$ is the loop group of $G$, and it was
argued that this generically carries a projective representation of $LG$, i.e. a representation of
the affine algebra $\hat{\mathfrak{g}}$ where $\mathfrak{g}=Lie(G)$. On a side note we remark that
the fact that the physical degrees of freedom are parametrised by $LG/G$, and not by the full $LG$,
is a close three-dimensional non-Abelian analogue of the redundant degrees of freedom present in the
boundary fields encountered in section~\ref{sec:canonical}. In the Abelian case this implies that
the chiral boson on the boundary does not contain zero-modes, i.e. that it can be written $\p\phi$
for some two-dimensional scalar field $\phi$.

A naive extrapolation to seven dimensions would imply that we should consider
$\M_7\cong \RR\times\Si$ where $\p\Si =\emptyset$. Furthermore we should allow for "marked surfaces"
on $\Si$. These are the analogues of marked points on $\Si_2$ since the closest 6D analogue of
chiral vertex operators are Wilson surfaces of chiral two-forms. Also in close analogy with 3D
Chern-Simons theory the marked surfaces should be thought of as intersections of Wilson three-
surfaces in $\M_7$ with $\Si$. To each such extended six-manifold should be assigned a vector space~
\footnote{although the more abstract formulation rather suggests other algebraic structures}, the
space of conformal blocks on $\Si$.
From this (admittedly very naive) picture it seems that the natural setting to formulate the
relation between 7D TFT and 6D chiral CFT is not in terms of local field theory, but rather on 2-
loop spaces. Very little is known concerning how to do this, although loop equations for the 6D
theories addressed here have been written down~\cite{Ganor:tlstr}.
For the 5D TFT it might not be completely unfeasible, however, since plenty of work has been done on
the loop formulation of Yang-Mills theory.
These analogies indicates strongly, we feel, that (higher-) loop variables are suitable to formulate
CFT's in diverse dimensions, it is therefore an issue we plan on returning to.

Let us finally note that there are many concepts which seem to be related to the type of theories
treated in the present work, but which are not yet very well understood. The results of~\cite{
Witten:qftjones} have been further developed into a mathematically rigorous formulation of three-
dimensional topological field theory, or rather several closely related formulations. There is no
unique definition, and although the original definition, to the best knowledge of the author, was
given by Atiyah~\cite{Atiyah:tft} it seems the formulation of Turaev~\cite{Turaev:knots} is at
present the most frequently adopted. The rigorous formulations involve quite abstract structures
such as modular tensor categories.
Regarding higher dimensional TFT's it has been suggested that n-categories, n-cobordisms and operads
are relevant. For discussions we refer to~\cite{BaezDolan:algtft} and references therein.
Higher p-form gauge fields with integer periods, present both in the TFT's and the chiral CFT's, are
presumably best understood in terms of {\em gerbes}~\cite{Giraud:nonabcoh, Brylinski:loopsp}. A
recent construction due to Chatterjee and Hitchin~\cite{Chatterjee:gerbs, Hitchin:slags} has made
Abelian gerbes, corresponding to free theories, accessible. The crucial objects are of course non-
Abelian gerbes which unfortunately so far lack a practical formulation. There are, however, recent
attempts~\cite{BreenMessing:dgerbes} to generalize the construction of~\cite{Chatterjee:gerbs,
Hitchin:slags} to non-Abelian gerbes so perhaps it will soon be possible to use this in constructing
interacting theories. There are also other ways to attack aspects of interacting theories in six
dimensions~\cite{Henningson:6dcft, Henningson:comrels, Gustavsson:tensorstrings} which are in some
sense more physical, and for that reason perhaps has a greater chance of success.
\vs{5mm}\\

\noi
{\bf Acknowledgment:} We would like to thank Stephen Hwang for valuable comments.
\vs{5mm}

\noi
{\bf Notes added:} After submission we have learned that the main results are already published by
Maldacena, Moore and Seiberg~\cite{MMS}. Different methods are used to obtain the results, however,
and the present paper may therefore still be of some interest.
The results in~\cite{MMS} are obtained using a covariant approach which certainly makes some aspects
more transparent, but questions regarding gauge invariance are more easily addressed in a canonical
formalism.
In the present paper the result is a consequence of demanding gauge invariance (or rather BRST
invariance) of the TFT on a manifold with boundary. It is completely independent of any form
of boundary conditions, in contrast, we believe, to the result of~\cite{MMS}.


\begin{thebibliography}{AA}
\bibitem{Witten:comments} E.~Witten {\em Some comments on string dynamics}, Los Angeles 1995, Future
perspectives in string theory 501-523, {\tt hep-th/9507121}
\bibitem{Strominger:openp} A.~Strominger {\em Open p-branes}, Phys.~Lett.~{\bf B383} (1996) 44-47, {
\tt hep-th/9512059}
\bibitem{Henningson:6dcft} M.~Henningson {\em A class of six-dimensional conformal field theories},
Phys.~Rev.~Lett.~{\bf 85} (2000) 5280, {\tt hep-th/0006231}
\bibitem{Witten:5brane} E.~Witten {\em Five-brane effective action in M-theory}, J.~Geom.~Phys.~{\bf
22} (1997) 103-133, {\tt hep-th/9610234}
\bibitem{HeNiSa:holfac} M.~Henningson, B.~E.~W.~Nilsson and P.~Salomonson {\em Holomorphic
factorization of correlation functions in $(4k+2)$-dimensional $(2k)$-form gauge theory}, JHEP {\bf
9909} (1999) 08, {\tt hep-th/9908107}
\bibitem{Verlinde:gdual} E.~Verlinde {\em Global aspects of electric-magnetic duality}, Nucl.~Phys.~
{\bf B455} (1995) 211-228, {\tt hep-th/9506011}
\bibitem{Schwarz:pnfn} A.~S.~Schwarz {\em The partition function of degenerate quadratic functional
and Ray-Singer invariants}, Lett.~Math.~Phys.~{\bf 2} (1978) 247-252
\bibitem{Witten:adstft} E.~Witten {\em AdS/CFT correspondence and topological field theory}, JHEP {
\bf 9812} (1998) 025, {\tt hep-th/9812012}
\bibitem{Witten:qftjones} E. Witten {\em Quantum Field Theory And The Jones Polynomial}, Commun.~
Math.~Phys.~{\bf 121} (1989) 351
\bibitem{EMSS:canqCS} S. Elitzur, G. Moore, A. Schwimmer and N. Seiberg {\em Remarks on the
canonical quantization of the Chern-Simons-Witten theory}, Nucl.~Phys.~{\bf B326} (1989) 108
\bibitem{MS:confzoo} G. Moore and N. Seiberg {\em Taming the conformal zoo}, Phys.~Lett.~{\bf B220}
(1989) 422
\bibitem{FjHw:99} J.~Fjelstad and S.~Hwang {\em Equivalence of Chern-Simons gauge theory and WZNW
model using a BRST symmetry}, Phys.~Lett.~{\bf B466} (1999) 227-233, {\tt hep-th/9906123}
\bibitem{Henningson:comrels} M.~Henningson {\em Commutation relations for surface operators in six-
dimensional $(2,0)$ theory}, JHEP~{\bf 0103} (2001) 011, {\tt hep-th/0012070}
\bibitem{KerlerLyuba:3Dcorner} T.~Kerler and V.~V.~Lyubashenko {\em Non-semisimple topological
quantum field theories for 3-manifolds with corners}, Springer (2001), Lecture notes in mathematics
1765, ISBN~3-540-42416-4
\bibitem{Ganor:tlstr} O.~J.~Ganor {\em Six-dimensional tensionless string theories in the large N
limit}, Nucl.~Phys.~{\bf B489} (1997) 95-121, {\tt hep-th/9605201}
\bibitem{Atiyah:tft} M.~F.~Atiyah {\em Topological quantum field theory}, Publ.~Math.~IHES~{\bf 68}
(1988) 175
\bibitem{Turaev:knots} V.~G.~Turaev {\em Quantum invariants of knots and three manifolds}, De
Gruyter (1994)
\bibitem{BaezDolan:algtft} J.~Baez and J.~Dolan {\em Higher-dimensional algebra and topological
quantum field theory}, J.~Math.~Phys.~{\bf 36} (1995) 6073--6105, {\tt q-alg/9503002}
\bibitem{Giraud:nonabcoh} J.~Giraud {\em Cohomologie non-Ab\'{e}lienne}, Grundlehren {\bf 179},
Springer Verlag, Berlin (1971)
\bibitem{Brylinski:loopsp} J.-L.~Brylinski {\em Loop spaces, characteristic classes and geometric
quantization}, Progress in Mathematics {\bf 107}, Birkh\"{a}user (1993), ISBN 0-8176-3644-7
\bibitem{Chatterjee:gerbs} D.~S.~Chatterjee {\em On the construction of abelian gerbs}, PhD thesis,
Cambridge (1998)
\bibitem{Hitchin:slags} N.~Hitchin {\em Lectures on special Lagrangian submanifolds}, {\tt math.DG/
9907034}
\bibitem{BreenMessing:dgerbes} L.~Breen and W.~Messing {\em Differential geometry of gerbes}, {\tt
math.AG/0106083}
\bibitem{Gustavsson:tensorstrings} A.~Gustavsson {\em The $d=6$, $(2,0)$-tensor multiplet coupled to
self-dual strings}, (2001) {\tt hep-th/0110248}
\bibitem{MMS} J.~Maldacena, G.~Moore and N.~Seiberg {\em D-brane Charges in Five-brane backgrounds},
JHEP~{\bf 0110} (2001) 005, {\tt hep-th/0108152}
\end{thebibliography}
\end{document}